\documentclass[journal=jacsat,manuscript=article]{achemso}
\usepackage{colortbl} 
\usepackage{graphicx} 

\usepackage{chemformula} 
\usepackage[T1]{fontenc} 
\usepackage{textcomp}
\setkeys{acs}{doi = true}
\setkeys{acs}{maxauthors=25}
\usepackage{fix-cm}
\usepackage{amsmath}
\usepackage{amsfonts}
\usepackage{amssymb}
\usepackage[table]{xcolor}
\usepackage{graphicx}
\usepackage{geometry}
\usepackage{float}
\usepackage{graphicx}
\usepackage{wrapfig}
\usepackage{times}
\usepackage{ragged2e}
\usepackage{upgreek}
\usepackage{xr}

\makeatletter
 \newcommand*{\addFileDependency}[1]{
   \typeout{(#1)}
   \@addtofilelist{#1}
   \IfFileExists{#1}{}{\typeout{No file #1.}}
 }
 \makeatother
\author{Yangchen He}
\affiliation[University of Wisconsin-Madison MS\&E]
{Department of Materials Science and Engineering,  University of Wisconsin-Madison, Madison, WI 53706, United States}

\author{Jessica Kienbaum}
\affiliation[University of Wisconsin-Madison MS\&E]
{Department of Materials Science and Engineering,  University of Wisconsin-Madison, Madison, WI 53706, United States}

\author{Wuzhang Fang}
\affiliation[University of Wisconsin-Madison MS\&E]
{Department of Materials Science and Engineering,  University of Wisconsin-Madison, Madison, WI 53706, United States}

\author{Hongrui Ma}
\affiliation[University of Wisconsin-Madison ECE]
{Department of Electrical \& Computer Engineering,  University of Wisconsin-Madison, Madison, WI 53706, United States}

\author{Ying Wang}
\affiliation[University of Wisconsin-Madison MS\&E]
{Department of Materials Science and Engineering,  University of Wisconsin-Madison, Madison, WI 53706, United States}
\alsoaffiliation[University of Wisconsin-Madison ECE]
{Department of Electrical \& Computer Engineering,  University of Wisconsin-Madison, Madison, WI 53706, United States}
\alsoaffiliation[UW-Madison Physics]
{Department of Physics, University of Wisconsin-Madison, Madison, WI 53706, United States}

\author{Ping Yuan}
\affiliation[University of Wisconsin-Madison MS\&E]
{Department of Materials Science and Engineering,  University of Wisconsin-Madison, Madison, WI 53706, United States}
\alsoaffiliation[UW-Madison Physics]
{Department of Physics, University of Wisconsin-Madison, Madison, WI 53706, United States}
\alsoaffiliation[UW-Madison Physics]
{Department of Chemistry, University of Wisconsin-Madison, Madison, WI 53706, United States}

\author{Daniel A. Rhodes}
\email{darhodes@wisc.edu}
\affiliation[University of Wisconsin-Madison MS\&E]
{Department of Materials Science and Engineering,  University of Wisconsin-Madison, Madison, WI 53706, United States}
\alsoaffiliation[UW-Madison Physics]
{Department of Physics, University of Wisconsin-Madison, Madison, WI 53706, United States}

\title[An \textsf{achemso} demo]
  {Achieving Large Uniaxial and Homogeneous Strain in Two-Dimensional Materials}

\begin{document}
\begin{tocentry}





    \includegraphics[width=1\linewidth]{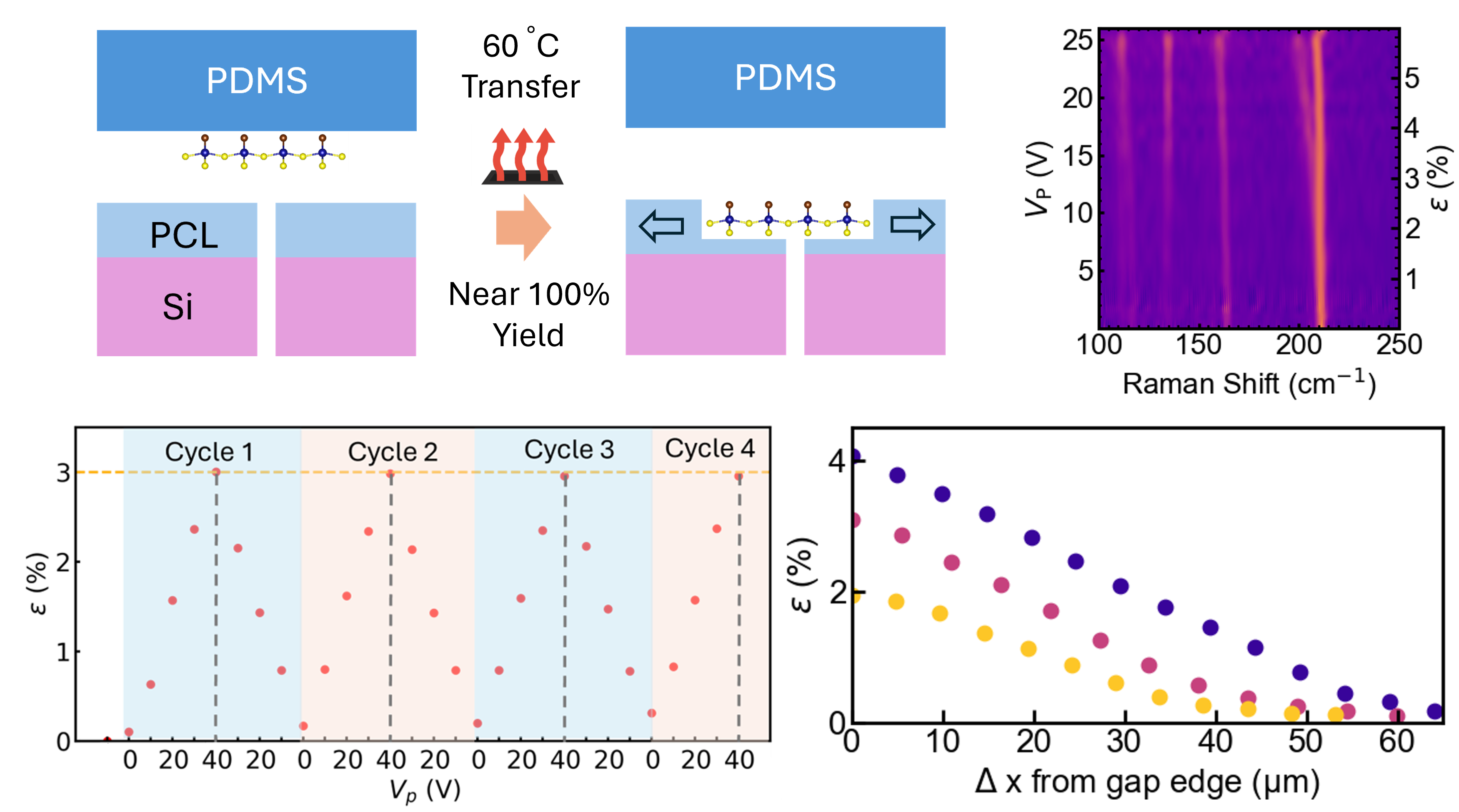}

 \end{tocentry}
\newpage
\begin{abstract}
\noindent Strain engineering is a powerful tool for tuning the electronic, magnetic, and topological properties of two-dimensional (2D) materials and thin films - particularly at high values of strain ($>3\%$) where many electronic, magnetic, and structural transitions have been predicted. However, most approaches to tuning strain in 2D materials are limited below 1.5\%, with poor repeatability when cycling strain and low strain transfer when cooling to cryogenic temperatures. Here, we report a high-yield sample preparation and device strain platform that overcomes these limitations, enabling precise, reversible strain tuning up to the intrinsic strain-to-failure of the materials tested herein. In addition, we show that this platform can be used to controllably design uniform linear strain gradients across  of 10's of $\upmu$m, providing a novel route to systematically investigate flexoelectric and flexomagnetic phenomena. Using CrSBr as a model system, we demonstrate uniform uniaxial strain, up to $\sim$4\%, with negligible slippage and linear strain gradients of up to 0.06\%/$\upmu$m. We further show that our strain approach is applicable to a broad class of 2D materials, validating its performance for three different phases of transition metal dichalcogenides: 2H-MoTe$_2$, 1T$^\prime$-MoTe$2$ and T$_\mathrm{d}$-WTe$_2$. In T$_\mathrm{d}$-WTe$_2$, verified by theoretical calculations, we show a continuous redshift of the A$_1^3$ mode, up to a record-breaking $\sim$5.5\% strain, with a clear separation of the A$_1^3$ and A$_1^2$ modes starting at 2\% strain. 
\end{abstract}

Strain in crystalline materials can strongly modify the electronic band structure and carrier interactions, enabling controlled tuning of their electronic~\cite{fischetti1996band,zeng2021tuning}, optical~\cite{chang2023impact}, and magnetic properties~\cite{shen2024strong}. Even modest values of strain ($<$1\%) can impact sensitive materials properties, such as superconductivity~\cite{steppke2017strong,jerzembeck2022superconductivity,malinowski2020suppression} and correlated electronic states~\cite{kazmierczak2021strain}. Two-dimensional (2D) materials, with elastic strains up to or exceeding 10\% before mechanical failure~\cite{bertolazzi2011stretching,cooper2013nonlinear}, offer a unique opportunity to explore the impact of extreme strain on material properties and concomitant functionalities, such as strain-engineering of carrier mobilities~\cite{datye2022strain} and optical sensing~\cite{gant2019strain}. Despite the multitude of advantages that large strain can offer, the integration of large strain as a reliable and widely tunable parameter in 2D materials and their heterostructures has remained technically challenging, with the majority of strain setups limited to $<$1.5\%.~\cite{iff2019strain,hou2019strain,cenker2022reversible,hu2023moire,michail2024biaxial,gao2025straining,harm2025strain}\\
\indent In recent years, there has been a substantial push to achieve highly tunable and in-situ control over uniaxial strain in 2D materials. However, these approaches have remained limited by low strain range, strain inhomogeneity, and poor reproducibility. In addition, in many of the approaches the reported strain was inferred from the substrate deformation (often with a transfer/gauge factor\cite{zhu2013strain}). However, strain of the 2D material can deviate from the nominal substrate strain due to poor strain transfer, for example, from slipping~\cite{li2020efficient,carrascoso2023improved} and nonuniform strain fields, particularly near anchored regions~\cite{liu2018approaching,harm2025strain}, necessitating careful mapping of the 2D materials properties to accurately evaluate the effectiveness of a given strain approach. One of the most common approaches is the use of a polymer substrate or encapsulant combined with a mechanical bending to transfer strain to the 2D material~\cite{ni2008uniaxial,zhu2013strain,desai2014strain,wang2019situ,li2020efficient}. The maximum strain values of bending geometries utilizing polymer substrates are limited by the 2D materials/polymer adhesion and can generate unintentional strain gradients and concomitant disorder from the underlying polymer/2D material interface~\cite{young2011strain,anagnostopoulos2015stress,androulidakis2019stress}.\\ 
\indent A potential solution to achieving large and uniform tensile strain is suspension of the 2D material across a narrow gap~\cite{zhang2014fracture,cao2020elastic}. A common strategy to suspend and strain the 2D material is to create a gap in a Si/SiO$_2$ chip (via propagation of a crack with a diamond scribe) mounted on a piezo stack and subsequently transfer the 2D material of interest over the gap~\cite{cenker2022reversible}. However, this approach is hindered by low yield due to the uncontrolled gap width, mismatch between heights of opposing gap edges, and limited adhesion between SiO$_2$ and the 2D material~\cite{Razorbill}. Although strains of up to 1.5\% have been reported using this method~\cite{cenker2022reversible}, the sample yield and achievable strain range are often inconsistent and strongly limited by the flake length along the strain axis (see Fig. S1). These issues narrow the selection of sample thickness and sample/substrate overlap to suppress slippage, further constraining sample choice and efficiency. To further solve these issues, flexible polymer substrates have been combined with anisotropic 2D substrates to effectively transfer strains of up to 2.3\% to 2D materials~\cite{cenker2025engineering}. This approach comes with the same strain inhomogeneity issues of earlier polymer substrate approaches, adds to fabrication complexity via the additional transfer steps, and remains limited to achievable strains tolerated by the underlying 2D substrate. In addition, the strong dependence of strain transfer on 2D material thickness, due to competition between interfacial friction and stress, still persists - with more than a $\sim$50\% reduction in material strain for thicker flakes and heterostructures with the same substrate deformation~\cite{cenker2025engineering}. To fully harness the potential of strain engineering in 2D materials, a more efficient and simpler approach that enables strains limited only by the intrinsic material properties, rather than by external disorder or sample dimension, needs to be developed.\\
\indent Here, we report a high-yield method for preparing uniaxial, tensile-strained 2D materials that reaches the intrinsic strain to failure limit of several 2D materials tested herein. Using this method, we demonstrate negligible slippage during cycling of strain with homogeneous strain across the majority of the intentionally strained area. This breakthrough in applying strain to 2D materials is achieved via three main approaches: first, creating a highly controllable gap and substrate alignment by incorporating lithographic and deep etching processes; second, by establishing a substrate cleaving procedure that maintains substrate height uniformity, therefore ensuring good contact during transferring and straining; and third, by enhancing strain transmission via functionalization of the Si/SiO$_2$ substrate with polycaprolactone (PCL).\\
\indent We validate the strain performance of our approach using 2D flakes of CrSBr, which has a well-calibrated Raman shift with uniaxial strain~\cite{cenker2022reversible}. Across multiple CrSBr flakes, we demonstrate a wide and tunable strain range with a strain to failure of over 3\% with temperatures down to 2 K. Multiple cycles of strain loading and unloading demonstrate effective strain modulation, revealing a reproducible high strain with a linear response with respect to displacement. These results are reproducible across multiple samples and materials with thicknesses of up to 100 nm. Large strain across a broad range of materials overcomes previous limitations\cite{cenker2022reversible,cenker2025engineering} and provides additional opportunities for strain engineering in 2D materials and heterostructures. Finally, we show that our platform enables homogeneous strain across the majority of the suspended sample surface. For regions supported by the underlying substrate, we show that our approach can be used to achieve tunable and homogeneous strain gradients, opening up opportunities to systematically investigate emergent phenomena such as flexomagnetism~\cite{qiu2023flexomagnetic,kitchaev2018phenomenology,eliseev2009spontaneous} and flexoelectricity~\cite{wang2019probing,zhuang2019intrinsic,tagantsev1985theory}. 

\section{Results and discussion}  
\begin{figure}[H]
    \includegraphics[width=\linewidth]{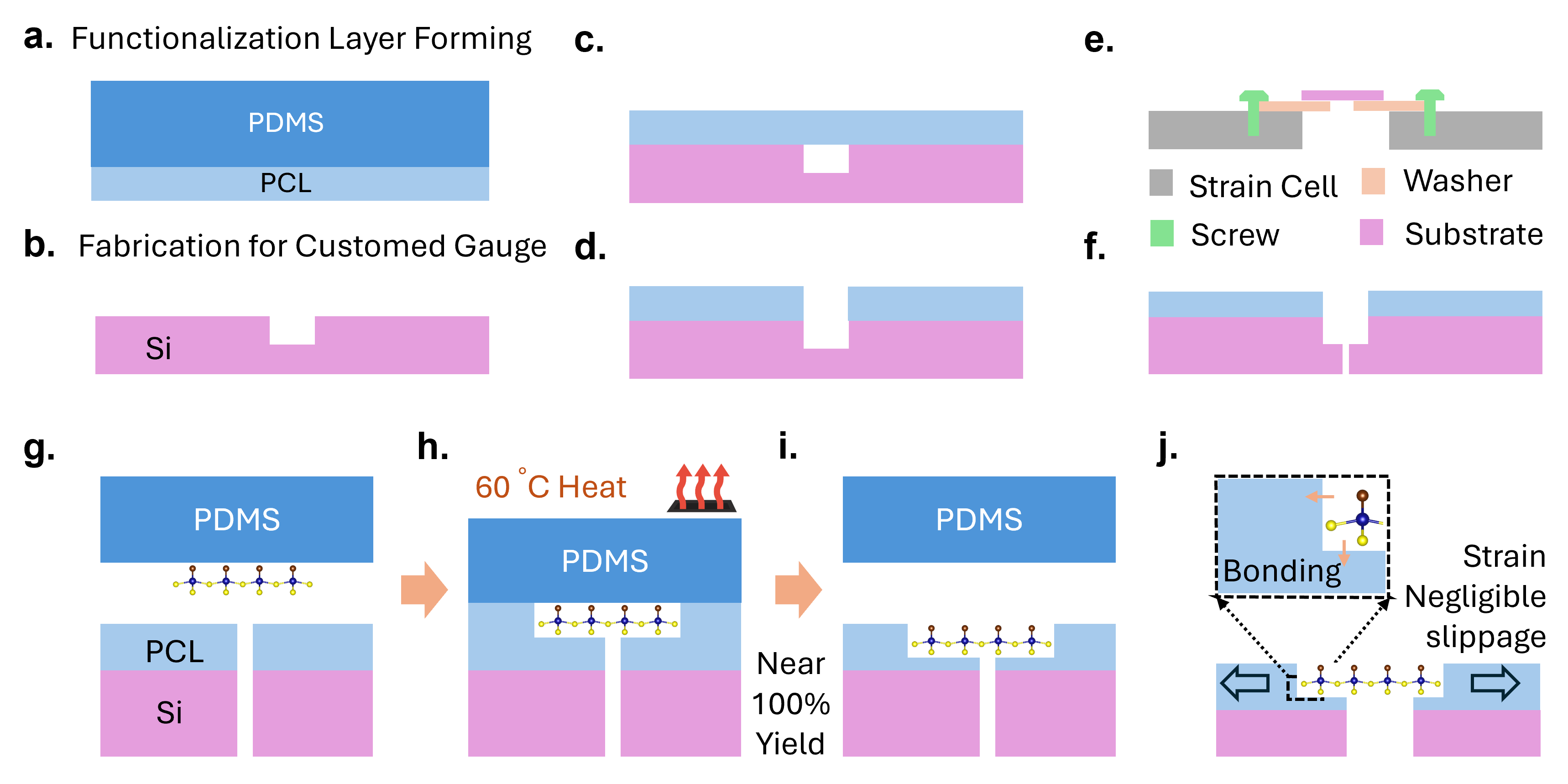}
    \caption{\justifying {Schematic for the strain substrate setup (a-e) and sample transfer process (g-i). 
    (j) Enlarged view of the PCL/2D material interface, where cooling through the glass transition temperature leads to encapsulation, enabling high-strain with negligible slippage.}
    }
\label{F1}
\end{figure}

\hspace{5mm} Fig. \ref{F1} depicts our approach to straining 2D materials utilizing a piezo-stack-based strain cell (Razorbill CS100) and a custom-fabricated substrate with an embedded trench. To design the gap over which we strain the 2D materials, we partially etched a gap (50 $\upmu$m) into a 100 $\upmu$m Si substrate using a Bosch process, where the width (5-50 $\upmu$m) of the gap defines the gauge length. Next, the surface of the Si substrate was functionalized with a PCL film via a PCL on polydimethylsiloxane (PDMS) transfer. We note that this functionalization layer is essential for achieving high transfer yield and strain transmission up to the point of failure. After fabrication and surface functionalization, the substrate is mounted by epoxy on the strain cell with a trench perpendicular to the strain direction. Finally, the monolithic substrate is cleaved into separated parts. We note that pre-defining the gap with a partial trench and cleaving of the substrate after mounting to the strain cell with epoxy is critical for maintaining the desired gap size upon cleaving and for achieving uniform height across the gap.\\
\indent Once the substrate is cleaved, the applied voltage directly determines the engineered gap through piezo-driven displacement of each side. To transfer the suspended 2D material across the gap, we use a dry transfer method\cite{wang2013one} with the substrate PCL acting as the pickup polymer. As demonstrated in by Fig. \ref{F1}g-i. we first prepare the sample on PDMS by directly exfoliating onto the PDMS for individual flakes or by pickup and flipping steps for heterostructures~\cite{kim2023imaging}. Subsequently, the PDMS/sample stack are centered with the gap and lowered to establish contact with the substrate. While maintaining substrate/stack contact, the temperature is raised above the glass transition temperature of PCL ($\sim$60 $^\circ$C), enabling encapsulation of the 2D material by the PCL and strengthening adhesion of the 2D material to the substrate. To avoid overheating of the piezos during this process, we use a hot air gun and heat the stack directly above the PDMS/sample stack. Finally, we cool the setup to room temperature and lift the PDMS, fully transferring the 2D material over the gap. Using this reversed dry transfer method, we achieve a nearly 100\% transfer yield of suspended samples.

\begin{figure}[H]
    \includegraphics[width=\linewidth]{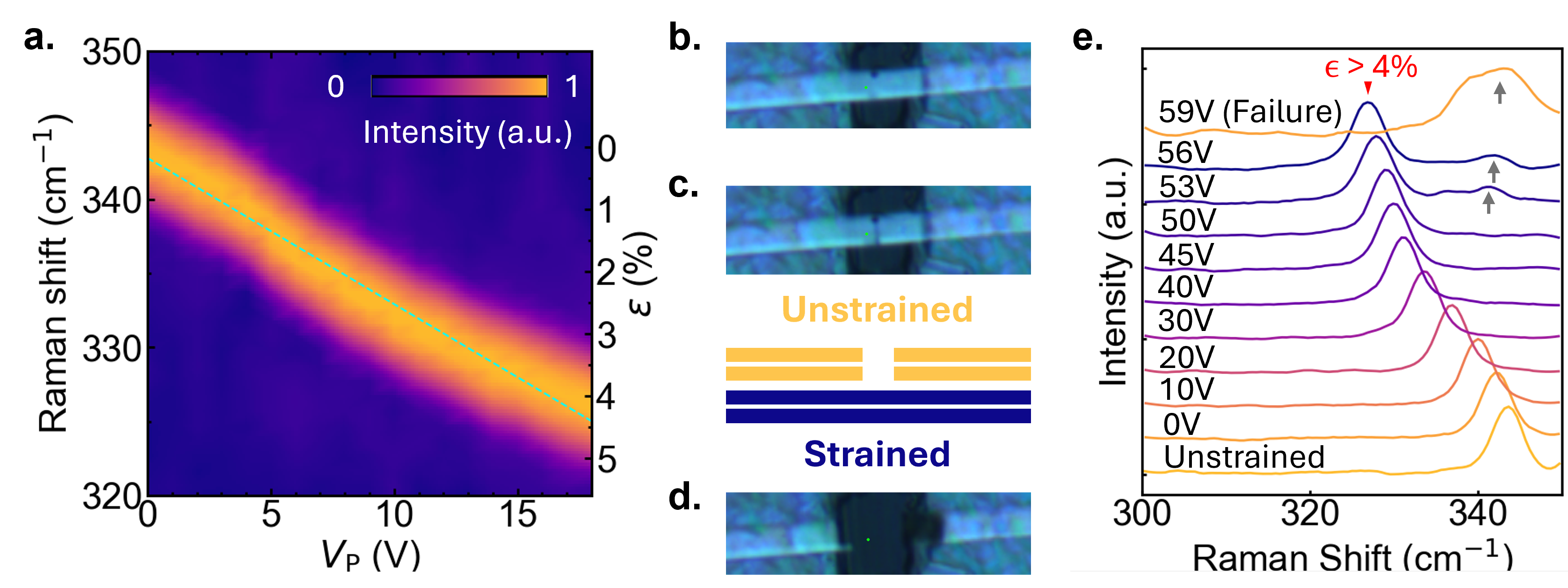}
    \caption{\justifying
    (a) Evolution of the Raman spectra centered near the A$^3_\mathrm{g}$ vibrational mode of CrSBr with increasing V$_\mathrm{p}$ in a thin flake (d $\approx$ 20nm). 
    (b-d) Optical images and schematics depicting the strain distribution of the CrSBr under homogeneous strain  (b), partial cracking (c) and mechanical failure (d). The optical images are taken after Raman spectra collection at 0 V, 50 V, 59 V, for c-e, respectively.
    (b) Evolution of the A$^3_\mathrm{g}$ mode with response to the increment of V$_\mathrm{p}$ towards mechanical failure (4$\%$) in a thick sample (d $\approx$ 40nm). Raman peak indicating relaxation status (grey arrow) at high strain and broadened peak at failure, indicating disordered strain status.
    }
\label{F2}
\end{figure}

\begin{figure}[H]
    \includegraphics[width=\linewidth]{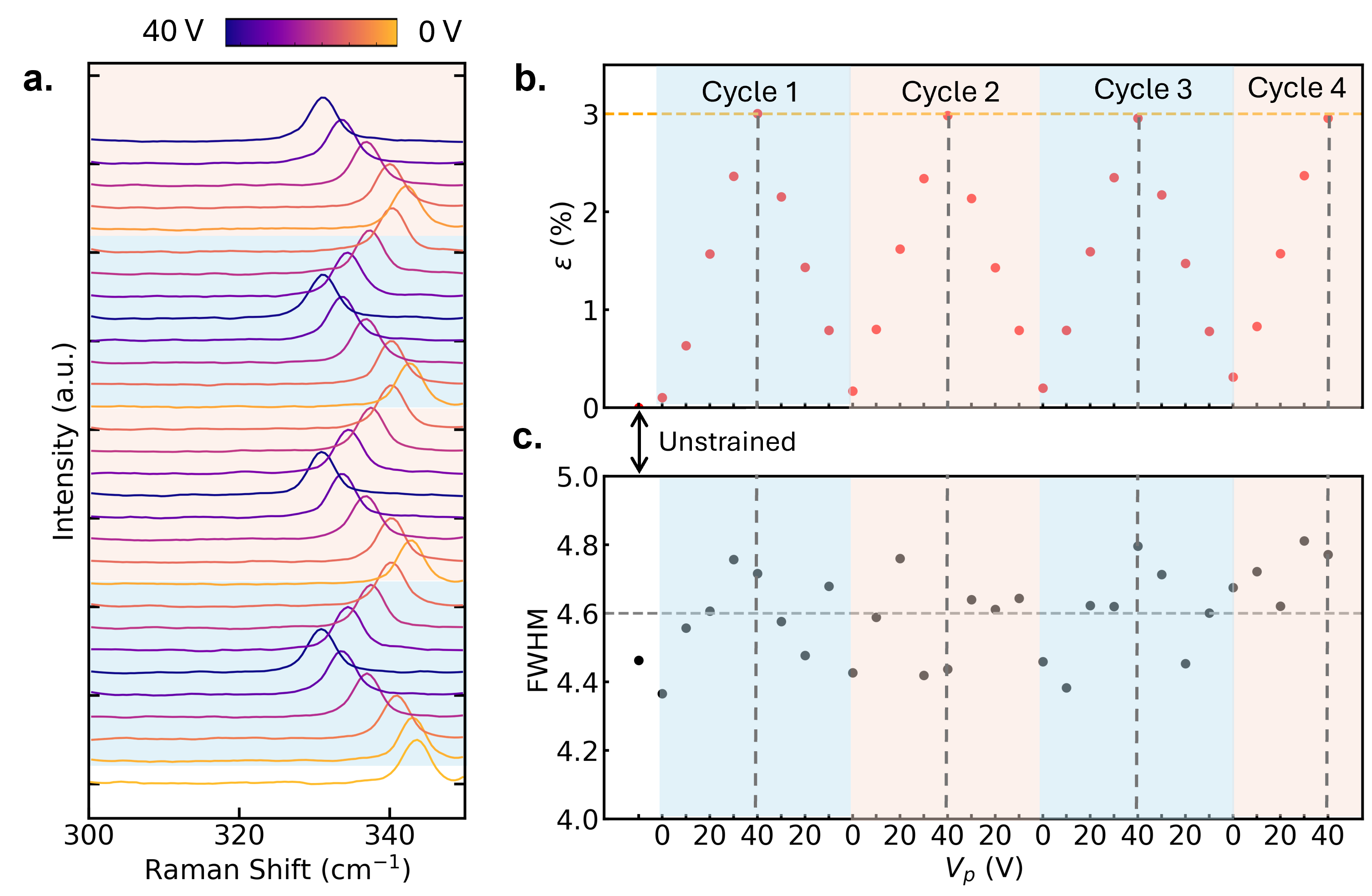}
    \caption{\justifying  
    (a) Raman spectra plotted with multiple cycles of V$_\mathrm{p}$ between unstrained and strained (3\%) states.
    (b) Converted strain from (a) across the four cycles.
    (c) FWHM of the corresponding A$^3_\mathrm{g}$ peak in (a). The grey dash vertical line marks the maximum V$_\mathrm{p}$ during the cycled test.
    }
\label{F3}
\end{figure}

To verify the efficient strain transfer and reproducible modulation of uniaxial strain of 2D materials toward mechanical failure of our process, we first apply this method to exfoliated flakes of CrSBr and measure the Raman response of the A$^3_\mathrm{g}$ mode as a function of voltage applied to the piezos (V$_\mathrm{p}$). We choose CrSBr as our model 2D material for demonstrating the efficiency of strain transfer, as it has been well calibrated in prior reports~\cite{cenker2022reversible}, where the A$^3_\mathrm{g}$ mode centered at 343.5 cm$^{-1}$ is sensitive to the strain status. From a thin flake (thickness, $d$, $\sim$20 nm) of CrSBr, we calibrated the strain response to the activation voltage V$_\mathrm{p}$. As shown in Fig. \ref{F2}a, increasing V$_\mathrm{p}$ leads to a monotonic red shift in the A$^3_\mathrm{g}$ mode up to a maximum of 17 cm$^{-1}$. Beyond this value, we see a physical cracking inside the suspended region. Using the previously calibrated strain-Raman shift rate of $\sim$ 4.2 cm$^{-1}$~\cite{cenker2022reversible,henriquez2025strain}, we converted the peak shift to the strain (represented as the right axis of Fig. \ref{F2}a). The calculated strain exhibits an approximately linear dependence with respect to V$_\mathrm{p}$, with a slope of 0.98\%/V. Although this strain sensitivity depends on specific device parameters such as the piezoelectric material type, temperature, and gauge length, the observed linear response suggests that V$_\mathrm{p}$ can serve as a reliable proxy for strain once the setup is calibrated.\\
\indent To demonstrate robust strain transfer of our setup and investigate thickness-dependent cracking behavior, we performed strain tests on a significantly thicker flake ($d \sim40$ nm). Due to the thermal expansion/shrinking of the PCL as it passes through the glass transition temperature during transfer, 2D materials are initially slightly strained in the suspended region. To zero out this offset, a small V$_\mathrm{p}$ is used to bring the sample back to 0\% strain, as seen in Fig. \ref{F2}e. In contrast to the abrupt failure observed in thinner flakes, the thick flake exhibited a multi-stage evolution: progressing from homogeneous elastic strain, to partial cracking, and finally to complete mechanical failure (Fig. \ref{F2}b–d). In the low-strain regime, the Raman spectra align with the trend of thinner CrSBr samples, showing a systematic red shift of the A$^3_\mathrm{g}$ peak (Fig. \ref{F2}e) and a linear strain response to V$_\mathrm{p}$ (Fig. S2). Probing the strain homogeneity in the CrSBr flake, we performed Gaussian fits of the A$^3_\mathrm{g}$ peak with increasing strain in Fig. \ref{F2}d. The summarized data in Fig. S2 shows negligible variation of the full width half maximum (FWHM), indicating that spatial homogeneity, within the laser spot size ($\sim$3 $\upmu$m$^2$), is preserved for strains of up to 3\% (V$_\mathrm{p}$ = 50 V). However, at $\epsilon > 3\%$ (50 V), we observe the emergence of a second peak near the value expected for the unstrained A$^3_\mathrm{g}$ mode, indicating relaxation of individual layers. This effect is likely caused by partial cracking, as seen from optical images of the CrSBr flake below and above 3.5\% strain (Fig. \ref{F2}b,c). The crack remains as the strain increase from 3\%, until clear mechanical failure (V$_\mathrm{p}$ = 59 V, $\epsilon > 4 \%$) is observed at the center of the gap, with the A$^3_\mathrm{g}$ mode shifting back to- and centered around the unstrained value and accompanied by a significant broadening of the FWHM from 4.6 to 10.8 cm$^{-1}$, indicating substantial relaxation from failure with a residual strain inhomogeneity (0-1\%). Concomitant with the blue shift in the Raman, optically we can see a clear breaking of the sample (Fig. \ref{F2}d), confirming the mechanical failure. \\
Beyond providing robust strain transfer for 2D materials at room temperature, to evaluate the cryogenic compatibility of our setup we cooled our CrSBr samples down to 2 K using the same sample preparation methods as described earlier, while maintaining an unstrained state during the cooling process. At 2 K, we incrementally increase V$_\mathrm{p}$ and monitored the Raman shift until visible failure of the thin film occurs. As shown in Fig. S3, both the thick ($d\sim$70 nm) and thin ($d\sim$25 nm) CrSBr flakes exhibit a linear redshifts of the A$^3_\mathrm{g}$ with increasing V$_\mathrm{p}$. Mimicking the behavior observed at room temperature, the thin CrSBr flake maintains a homogeneous strain state until mechanical failure. In contrast, the thicker CrSBr flake undergoes partial relaxation prior to failure, as evidenced by the emergence of a secondary, blue-shifted Raman peak aligned with the unstrained state. Both thin and thick flakes reach mechanical failure in our setup, with maximum detected strains of 3.16 and 2.16\% for thick and thin, respectively. Note that the strain to failure of a specific sample is governed not only by the material's intrinsic phonon instability\cite{cooper2013nonlinear,liu2007ab}, but also by extrinsic factors such as defects and dislocations\cite{zhang2014fracture,dang2014effect}. These extrinsic factors can lead to significant sample-to-sample variability. Crucially, our platform can support multiple samples within a single transfer process, significantly increasing the throughput of material screening and determination of intrinsic limits with strain.\\
\indent Having established the capability of reaching the max strain limit for CrSBr, we next evaluate the reliability of strain transfer under cyclic loading. To verify the absence of interfacial slippage and assess the reproducibility of our setup, we subjected the sample to multiple strain cycles, oscillating between 0\% and 3\%. These values are chosen to avoid partial cracking. As expected, from the Raman spectra across the multiple cycles, shown in Fig. \ref{F3}a, we do not observe any emergence of a secondary peak related to partial relaxation. Converting the Raman peak values to strain as a function of V$_\mathrm{p}$ (Fig. \ref{F3}b), we can see that strain values maximize and minimize repeatably at maxima and minima of V$_\mathrm{p}$. From the FWHM of the same Raman spectra, Fig. \ref{F3}c, we observe randomly distributed between 4.4 to 4.8 cm$^{-1}$, indicating no appreciable change. Within the gap, the relatively unchanged FWHM confirms the homogeneity of the strain transferred to the suspended CrSBr and strong anchoring by the PCL with negligible slippage under repeated loading. The stable, linear relationship between V$_\mathrm{p}$ and $\epsilon$ provides a built-in calibration: once this proportionality is established, the V$_\mathrm{p}$ itself can serve as an accurate indicator of strain during other experimental measurements where direct tracking of strain is not possible.

\begin{figure}[H]
    \includegraphics[width=\linewidth]{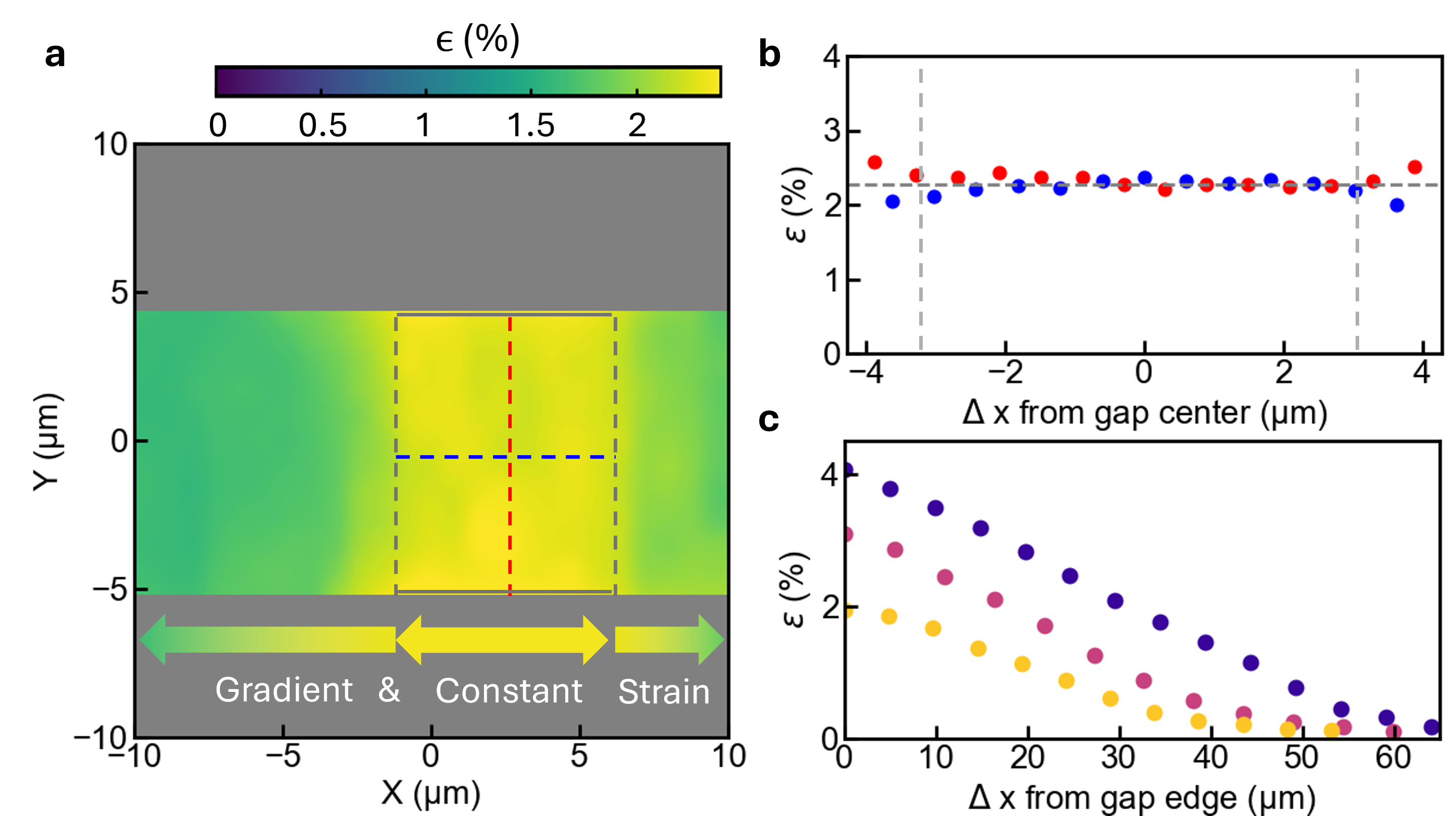}
    \caption{\justifying
    (a) Spatial map of tensile strain derived from the CrSBr A$_\mathrm{g}^\mathrm{3}$ Raman mode, showing homogeneous strain within the gap region (grey dashed line shows gap edge and solid lines indicate transverse film edge) and a strain gradient in the adjacent regions. (b) Strain line profiles parallel (blue) and perpendicular (red) to the strain direction, extracted from the constant strain region in (a).
    (d) Strain distribution extending outside the gap. By fixing the gap strain at specific levels (4\%, 3\%, 2\%), a linear strain gradient is maintained across different strain backgrounds.
    }
\label{F4}
\end{figure}

\indent Measuring the distribution of strain for our approach, we map the Raman spectra of suspended CrSBr flakes, held at 2\% strain, within and outside of the gap, shown in Fig. \ref{F4}a. Within the gap, the A$^3_\mathrm{g}$ mode exhibits a uniform redshift corresponding to a $\sim$2\% uniaxial strain, a consistency maintained regardless of whether the scan trajectory is parallel (Fig. \ref{F4}b, blue) or perpendicular (Fig. \ref{F4}b, red) to the uniaxial strain direction. Near the edges, the strain distribution is not perfectly homogeneous and slightly decreases (gap edge) or increases (transverse edge). These changes in strain are likely due to differences in edge constraints between anchored and unanchored regions, which can lead to enhanced effective strain at the free boundaries (i.e., transverse edge). Transitioning from the suspended to the anchored region, the strain begins to relax starting from near the gap edge (Fig. \ref{F4}b, vertical dash-line). Intuitively, one expects a strain gradient in the CrSBr flake extending from the freestanding region into the region clamped by the PCL, with a decay length determined by the adhesion between the PCL and CrSBr. As shown in Fig. S4a, 60 $\upmu$m line scans of the Raman spectra confirm this symmetric strain decay starting from the gap edges and into the clamped region. Higher resolution mapping extending into the clamped region (Fig. S4b) further verifies the linearity of this relaxation, yielding a monotonically 
decreasing, linear strain gradient of $\sim$ 0.07\% $\mu$m$^{-1}$ over 10's of $\upmu$m in length. These results showcase the dual capability of our strain approach, yielding large homogeneous areas of both high strain (suspended) and strain gradients (supported).\\ 
\indent In order to understand the limitations of achievable strain gradients with our approach, we further measured the supported region at fixed strains of 4, 3, and 2\% (Fig.\ref{F4}c). Linear fits of the relaxation regime (Fig. S5) reveal distinct strain gradients of 0.071\% $\upmu$m$^{-1}$, 0.061\% $\upmu$m$^{-1}$, 0.052\% $\upmu$m$^{-1}$, respectively. These results demonstrate the direct correlation between V$_\mathrm{p}$ and the resulting values of the strain gradient and showcase the control over tunability of these parameters at the $\upmu$m-scale. In addition to applied uniaxial strain, the values of the strain gradients and stress transfer lengths achieved above are likely determined by a combination of the interfacial shear stress at the van der Waals interface~\cite{androulidakis2019stress}, the elastic modulus of the PCL, and the boundaries given by the geometry of the CrSBr flake. In agreement with this picture, at constant V$_\mathrm{p}$ we find that the strain gradient can significantly differ depending on the length of CrSBr anchored by the PCL. For values below a critical length ($40-60$ $\upmu$m depending on V$_\mathrm{p}$, see Fig. \ref{F4}c), the CrSBr flakes prefer to distribute the strain across the entire anchored length, fully relaxing back to the V$_\mathrm{p}=0$ V strain value set by the interface with the PCL ($\sim0.25\%$). Figure S6 highlights this effect, where one side of a CrSBr flake formed a wrinkle, significantly shortening the anchor length to 9 $\upmu$m and enhancing the strain gradient as compared to the oppositely anchored end (38 $\upmu$m in length), which experiences a small strain gradient of $\sim$0.25\%. 

\begin{figure}[H]
    \includegraphics[width=\linewidth]{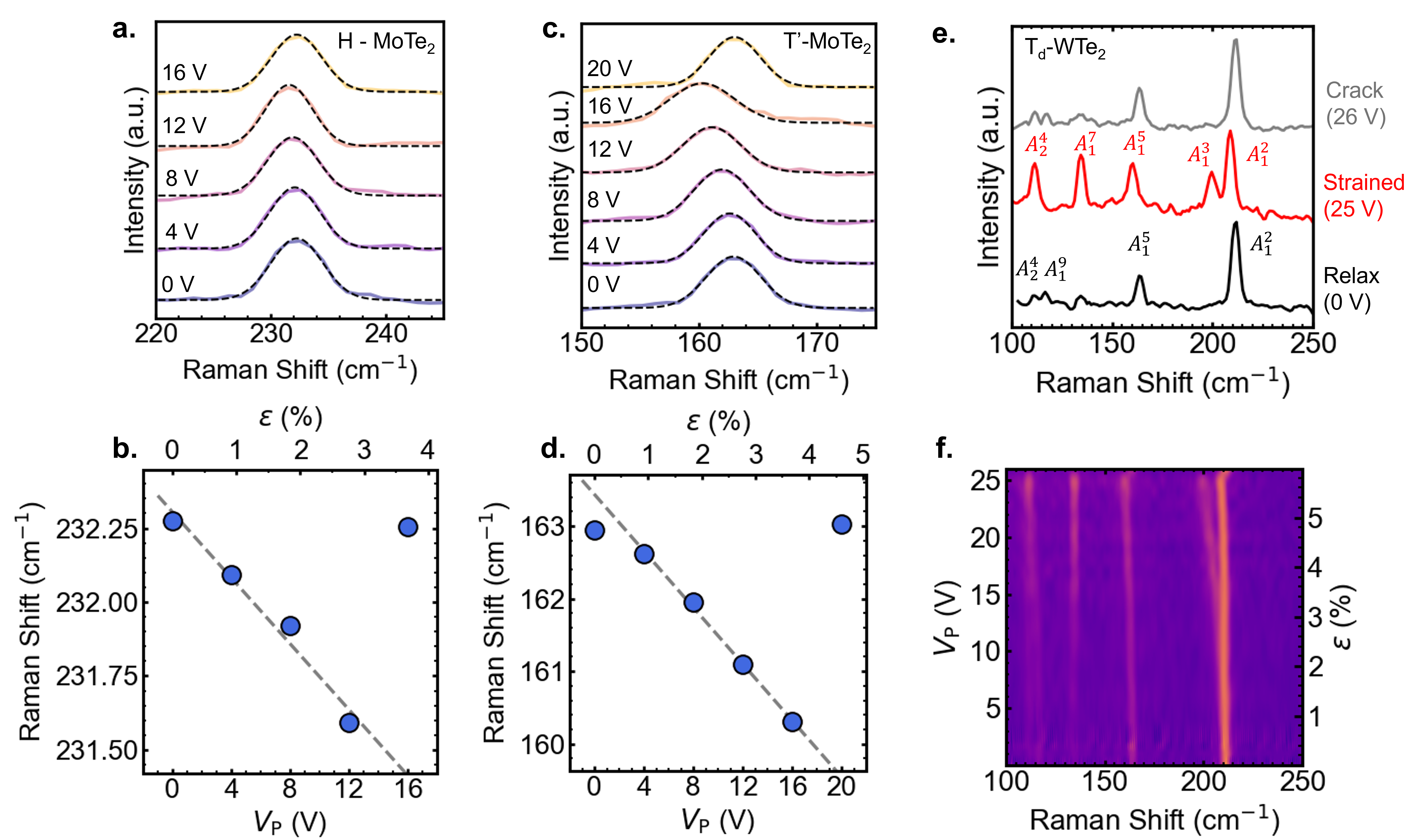}
    \caption{\justifying
    Evolution of Raman spectra for mechanically exfoliated (a) 2H-MoTe$_\mathrm{2}$, (c) 1T$^\prime$-MoTe$_\mathrm{2}$, and (e) T$_\mathrm{d}$-WTe$_\mathrm{2}$, under increasing piezoelectric voltage (V$_\mathrm{p}$). Dashed lines indicate Gaussian fits to the phonon modes exhibiting the strongest strain response. The linear evolution of the Raman peak position as a function of V$_\mathrm{p}$ is shown for (b) (2H-MoTe$_\mathrm{2}$) and (d) (1T$^\prime$-MoTe$_\mathrm{2}$). (f) Intensity map of T$_\mathrm{d}$-WTe$_2$ spectra versus V$_\mathrm{p}$, highlighting the strain-induced enhancement of the A$_{1}^{7}$ modes and the splitting of the A$_{1}^{2}$ mode. For figure (b, d, f), the corresponding tensile strain values (right axis) were calculated using a calibration factor of 0.23\% V$^{-1}$, derived from the CrSBr sensor on the same setup.
    }
\label{F5}
\end{figure}

Finally, we show that our strain approach can be applied to a broad range of 2D materials, with the maximum accessible strain limited by the material's intrinsic fracture strength. Highlighting this broad applicability, demonstrate broad applicability across different structural phases, we extended our platform three different TMD structures: hexagonal 2H-MoTe$_2$, monoclinic 1T$^\prime$-MoTe$_2$ and orthorhombic T$_\mathrm{d}$-WTe$_2$. Each of these materials are well known to be strongly tunable by strain, with changes that can be easily tracked by Raman~\cite{jo2019magnetoelastoresistance,xu2025dual,yip2023drastic,yang2019raman}. For each material tested, the strain-sensitive Raman modes exhibited a linear dependence on $V_\mathrm{p}$ up to the point of fracturing (Fig.\ref{F5}) - confirming efficient strain transfer and the ability to access strain values approaching the material's intrinsic limit. For 2H-MoTe$_2$ (Fig. \ref{F5}a,b) and 1T$^\prime$-MoTe$_2$ (Fig. \ref{F5}c,d), we find large strain to failure values of 2.5 and 3.5\%, respectively. In T$_\mathrm{d}$-WTe$_2$, we reach a significantly larger strain to failure value of 5.5\% (applied along the $a$ axis) . As shown in Fig. \ref{F5}e, for T$_\mathrm{d}$-WTe$_2$ we observe four dominant optical modes in the relaxed state: A$^4_2$ (110 cm$^{-1}$), A$^9_1$ (120 cm$^{-1}$), A$^5_1$ (163 cm$^{-1}$) and A$^2_1$ (211 cm$^{-1}$), consistent with previous reports~\cite{jiang2016raman}. Increasing strain causes the A$^4_2$ and A$^9_1$ modes to overlap at $V_\mathrm{p}=25$ V, resulting in a single, significantly enhanced peak intensity at 110 cm$^{-1}$, while a new mode at 135 cm$^{-1}$, assigned to A$^7_1$, gradually intensifies with increasing strain (Fig. \ref{F5}f). The A$^5_1$ mode exhibits monotonic redshift, with a slight intensification as strain increases. For the peak centered at 211 cm$^{-1}$ (A$^2_1$) redshifts linearly between 0-2\% strain. Above 2\% strain, this peak splits into two different Raman modes. Aside from the intensities, the evolution of these Raman modes are well captured by our density functional theory calculations (Fig. S7). The splitting of the A$^2_1$ peak at high strain can be attributed to the increased frequency separation of the optically active A$^3_1$ mode, where increased strain results in a linearly increasing separation of the A$^3_1$ and A$^2_1$ modes, going from just 2 cm$^{-1}$ in the relaxed state to 12 cm$^{-1}$ at 5\% strain. Experimentally, we observe just a $\sim$10 cm$^{-1}$ splitting of the A$^3_1$ and A$^2_1$ modes at 5\% strain, which we attribute to resolution limits of our setup. 
%
%
%

In summary, we report a high-yield device-fabrication method that enables efficient, deterministic strain transfer, is fully compatible with in-situ strain-tuning setups, and is applicable to a broad range of 2D materials. Using CrSBr as our model 2D material, we demonstrated uniform uniaxial strains of up to $\sim$4\%—limited only by sample fracture—with negligible slippage between cycles. Notably, in T$_\mathrm{d}$-WTe$_2$, we reached strains of 5.5\% before fracturing, the highest value to date. In addition to uniform strain, our approach establishes a controllable, linear strain gradient extending up to 40 $\upmu$m with a homogeneous spatial distribution. To our knowledge, this is the first such demonstration, opening up opportunities to systematically explore flexoelectricity, flexomagnetism, and other related phenomena. Our strain approach offers a significant advantages over conventional techniques. Unlike methods relying on substrate height modulation~\cite{palacios2017large}, stochastic gradients found in wrinkled films utilizing pull and release methods~\cite{du2021epitaxy,castellanos2013local}, or strain transfer from layers of different materials~\cite{cenker2025engineering} or on polymer substrates~\cite{cenker2025engineering, li2020efficient}, our approach maintains a relatively flat and planar geometry without any constraint on the maximum strain that can be applied. Altogether, our approach - as currently implemented - has high yield, with nearly 100\% transfer success, and enables large strains while also maintaining efficiency at cryogenic temperatures. Our approach can be readily combined with optical spectroscopy techniques and is easily adaptable for other measurement modalities, including synchrotron-based X-ray techniques, angle-resolved photoemission spectroscopy, and magnetotransport measurements. Ultimately, this platform provides a continuous, reversible route to explore strain-dependent electronic structure, band topology, and emergent quantum phases across the strain regime in a single material platform.

\begin{acknowledgement}
This work was completely supported by the University of Wisconsin Materials Research Science and Engineering Center (NSF DMR-2309000). J.K. acknowledges the Science, Mathematics, and Research for Transformation (SMART) Department of Defense scholarship program.The authors gratefully acknowledge the use of facilities and instrumentation in the Wisconsin Center for Nanoscale Technology. This center is partially supported by the Wisconsin Materials Research Science and Engineering Center (NSF DMR-2309000) and by the University of Wisconsin–Madison.
\end{acknowledgement}




\newpage
\bibliography{reference_Dec.bib}

\end{document}



\maketitle

\section{Methods}\vspace{-3mm}
\noindent \textbf{Crystal Synthesis:}\\ CrSBr single crystals were obtained from HQ Graphene, and the tellurides (T\textsubscript{d}-MoTe\textsubscript{2}, 2H-MoTe\textsubscript{2}, and WTe\textsubscript{2}) were synthesized by self-flux growth. The synthesis of T\textsubscript{d}-MoTe\textsubscript{2} followed the procedure described in our previous work.\cite{li2024twofold} To synthesize 2H-MoTe\textsubscript{2}, Mo powder (99.997\%) and Te lumps (99.9999\%) were combined in a molar ratio of 1:15 inside a Canfield crucible (LSP Ceramics), which was subsequently sealed under vacuum ($\sim 5 \times 10^{-6}$~Torr) in a quartz ampoule. The reagents were heated to 1120~$^\circ$C over 12~h, held at this temperature for 5~days, and then cooled to 525~$^\circ$C at a rate of 0.89~$^\circ$C/h. At 525~$^\circ$ C the excess Te flux was decanted in a centrifuge, quenching the crystals in air. The harvested 2H-MoTe\textsubscript{2} single crystals were sealed in a second evacuated quartz ampoule and annealed at 425~$^\circ$C with a 200~$^\circ$C gradient for 48~h to remove residual tellurium. To synthesize WTe\textsubscript{2}, W powder (99.999\%) and Te lumps (99.9999\%) were combined in a molar ratio of 1:25 and sealed as described above. The reagents were heated to 1120~$^\circ$C over 24~h and held at this temperature for 3~days. The temperature was then ramped down to 800~$^\circ$C over 72~h and ramped back to 1120~$^\circ$C over 5~h for two cycles, followed by slow cooling to 525~$^\circ$C at 0.89~$^\circ$C/h. At 525~$^\circ$C the excess Te flux was removed, and following the annealing process described above, WTe\textsubscript{2} single crystals were harvested.\vspace{2mm}\\
\noindent \textbf{Substrate fabrication:}\\ To fabricate a substrate with a deep trench of defined width, we start with a 100 $\upmu$m thick Si wafer and use a single-layer mask of AZ 12XT-20PL-10. A 10 $\upmu$m thick resist was formed by spin-coating. A Heldeiberg instrument DWL 66+ Laser Lithography System was used to pattern the photoresist and define the trench width. After exposure, the Si chip was developed and etched in an STS Deep Reactive Ion Si Etcher using a Bosch process. We used an Si etch rate of 1 $\upmu$m/min, with a total etch depth of $\sim$50 $\upmu$m.\vspace{2mm}\\
\noindent \textbf{Functionalization:}\\ The fabricated substrate was functionalized with a Polycaprolactone (PCL) film. First, a solution of PCL in tetrahydrofuran (THF) was prepared by directly adding and stirring 1 wt\% PCL into THF. The PCL/THF solution was then spin coated onto a polydimethylsiloxane (PDMS) substrate at 2000 rpm to produce a thin PCL film. Subsequently, the PDMS/PCL cube was diced and aligned with the trench and heated to 60 $^\circ$C to fully release and transfer the PCL onto the chip.\vspace{2mm}\\
\noindent \textbf{Mount Procedure:}\\ The functionalized substrate is mounted on the strain cell and then cracked to ensure negligible gap opening and a leveled surface of the Si chip for the sample (\textit{e.g.}, CrSBr, WTe$_2$, and MoTe$_2$) transfer. First, a pair of standard SL sample plate (Razorbill) were fixed on the strain cell using two screw-washer sets, and the fabricated substrate was epoxied on the sample plates with Stycast 2850FT and cured at 60 $^\circ$C for 3 hrs. The processing temperature is carefully controlled below 100 $^\circ$C to be compatible with the piezostack and electronics. After curing, a razor blade is used to apply force beneath the trench to cleave the chip into two along the deep trench, while mantaining the matched height and gap size. \vspace{2mm}\\
\noindent \textbf{Transfer Process:}\\ Samples were exfoliated onto PDMS and identified under optical microscope. Afterwards, the PDMS/sample stack was aligned over the trench in the Si chip. After making contact between PDMS/sample and substrate, the entire transfer stack was heated to and held at 60 $^\circ$C (above the glass transition temperature of PCL), for 20 s. Finally, the entire stack was cooled to room temperature and the PDMS was released from the sample/substrate. While transfers of 2D materials often fail when transferring over gaps, this process is a high-yield transfer process with zero failures over many attempts.\vspace{2mm}\\
\noindent \textbf{Raman Measurements:}\\ Room-temperature Raman spectra were measured using a Horiba LabRAM HR Evolution Confocal Raman Microscope with an excitation of 532 nm, filter of 1\% and power of 100 mW. Temperature-dependent Raman scattering measurements were performed in a backscattering geometry using a custom-built micro-Raman system integrated with a Quantum Design OptiCool cryostat. The sample was excited using a 633 nm laser, with the incident power maintained below 1 mW at the sample surface to minimize local laser-induced heating. The excitation beam was focused onto the sample using a long-working-distance 20$\times$ objective (NA = 0.41) with glass correction, yielding a focal spot size of approximately 2 $\upmu$m. The scattered signal was collected through the same objective and dispersed through an Andor Shamrock 500i spectrometer with an 1800 lines/mm grating. Spectra were recorded using a high-sensitivity Andor iDus CCD detector. Data acquisition was managed via Andor Solis software, providing the spectral resolution necessary to track evolution of the vibrational modes across the measured temperature range.\vspace{2mm}\\
\noindent \textbf{Strain-dependent DFT Calculation:}\\ Density functional theory calculations were performed using the Vienna Ab Initio Simulation Package (VASP)~\cite{VASP1,VASP2,VASP3} and the Quantum ESPRESSO (QE) code~\cite{QE1,QE2}. For VASP, the projector augmented wave (PAW) method~\cite{Blochl1994} was used with a kinetic energy cutoff of 500 eV. For QE, the optimized norm-conserving Vanderbilt (ONCV) pseudopotentials from the PseudoDojo project~\cite{ONCV1,ONCV2} were used with a kinetic energy cutoff of 90 Ry. The generalized gradient approximation of Perdew-Burke-Ernzerhof (PBE)~\cite{PBE} was used as the exchange-correlation functional. The DFT-D2 method was used to account for the Van de Waals interactions~\cite{grimme_consistent_2010}. A $\Gamma$-centered $12\times6\times3$ k-point mesh was used for integration in the first Brillouin zone. Structures were fully relaxed using VASP until the force on each atom was less than 5 meV/\AA. For WTe$_2$, the resulting relaxed lattice constants were a=3.541, b=6.244, and c=13.761 \AA, in good agreement with experiment~\cite{dawson1987electronic}. A uniaxial strain was applied along the $a$ axis. Phonon properties were calculated using density functional perturbation theory (DFPT) implemented in QE ~\cite{DFPT1,DFPT2}. 
WTe$_2$ has a point group of $C_{2v}$ and the energy and irreducible representation label at the $\Gamma$ point for 0\% strain were summarized in Table \ref{tab:modes_symmetry}. Based on the symmetry analysis, all phonon modes are Raman active. 

\begin{figure}[H]
    \centering
    \includegraphics[width=0.5\linewidth]{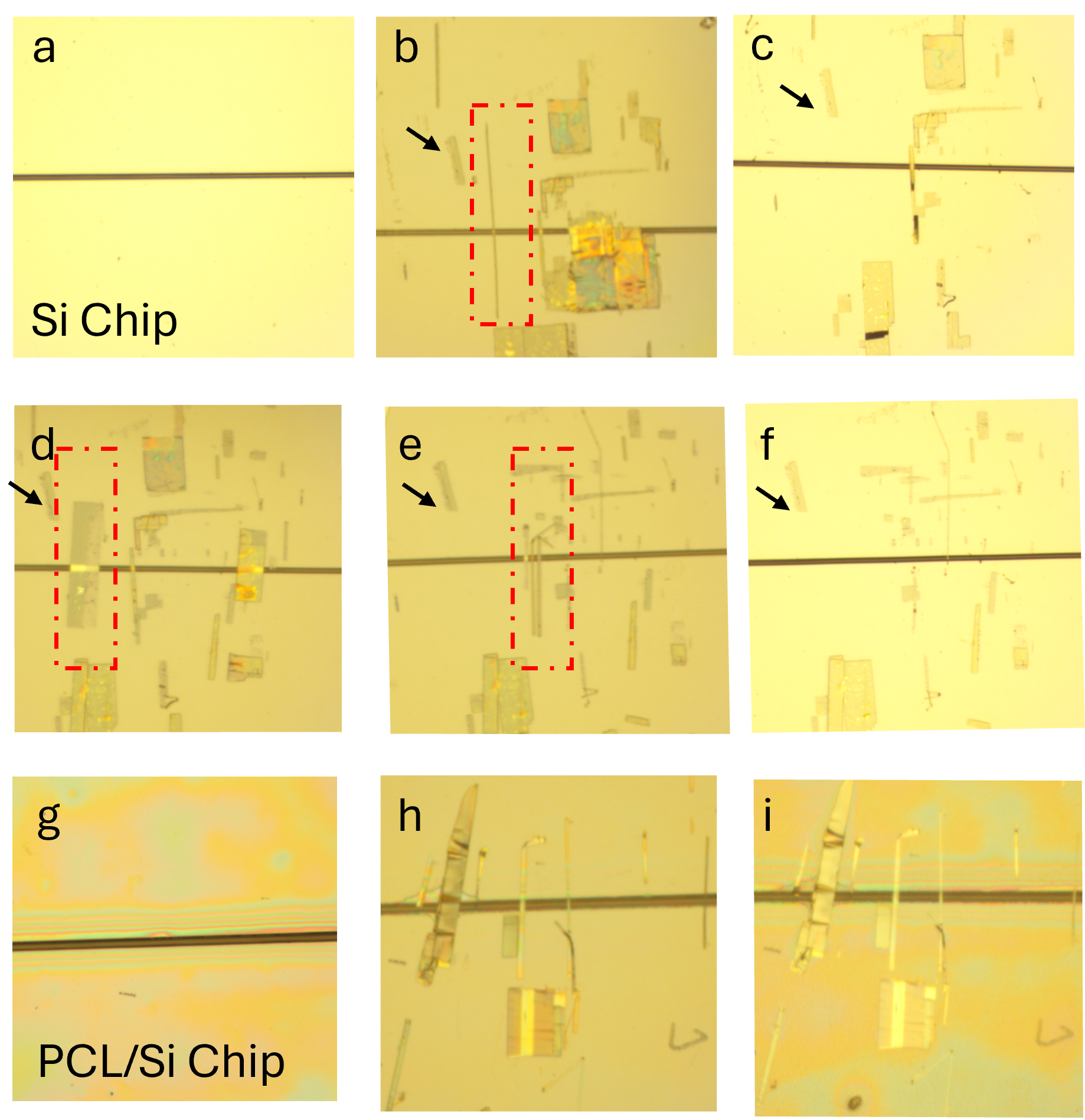}
    \caption{\textbf{Enhanced flake adhesion using PCL functionalization.} (a–f) Failed transfer attempts on a bare Si chip. (b) PDMS stamp with flakes brought into contact with a trench in an Si chip. (c) Result after stamp release, showing poor adhesion. (d–f) Two subsequent transfer attempts on the same location showing similar failure, indicating insufficient adhesion between the target samples and the bare Si surface. (g–i) Successful transfer on a PCL-functionalized Si chip with a trench. (g) Optical image of the PCL-functionalized Si surface. (h) PDMS stamp in contact and (i) after release. The flakes strongly adhere to the PCL, enabling high-yield suspension across the trench.}
\label{Si_slippage}
\end{figure}


\begin{figure}[H]
    \centering
    \includegraphics[width=0.5\linewidth]{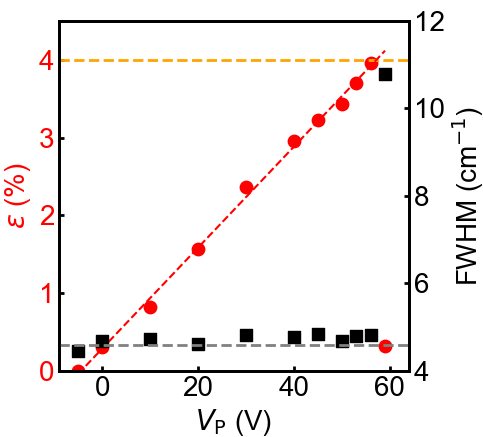}
    \caption{\textbf{Calibration of strain versus applied piezoelectric voltage ($V_\textrm{P}$). }Strain values and full width half maximums (FWHM) were derived from Gaussian fits of the Raman spectra (Fig. 2e in main text) and show a linear response to $V_\textrm{P}$.}
\label{thicklinear}
\end{figure}

\begin{figure}[H]
    \centering
    \includegraphics[width=\linewidth]{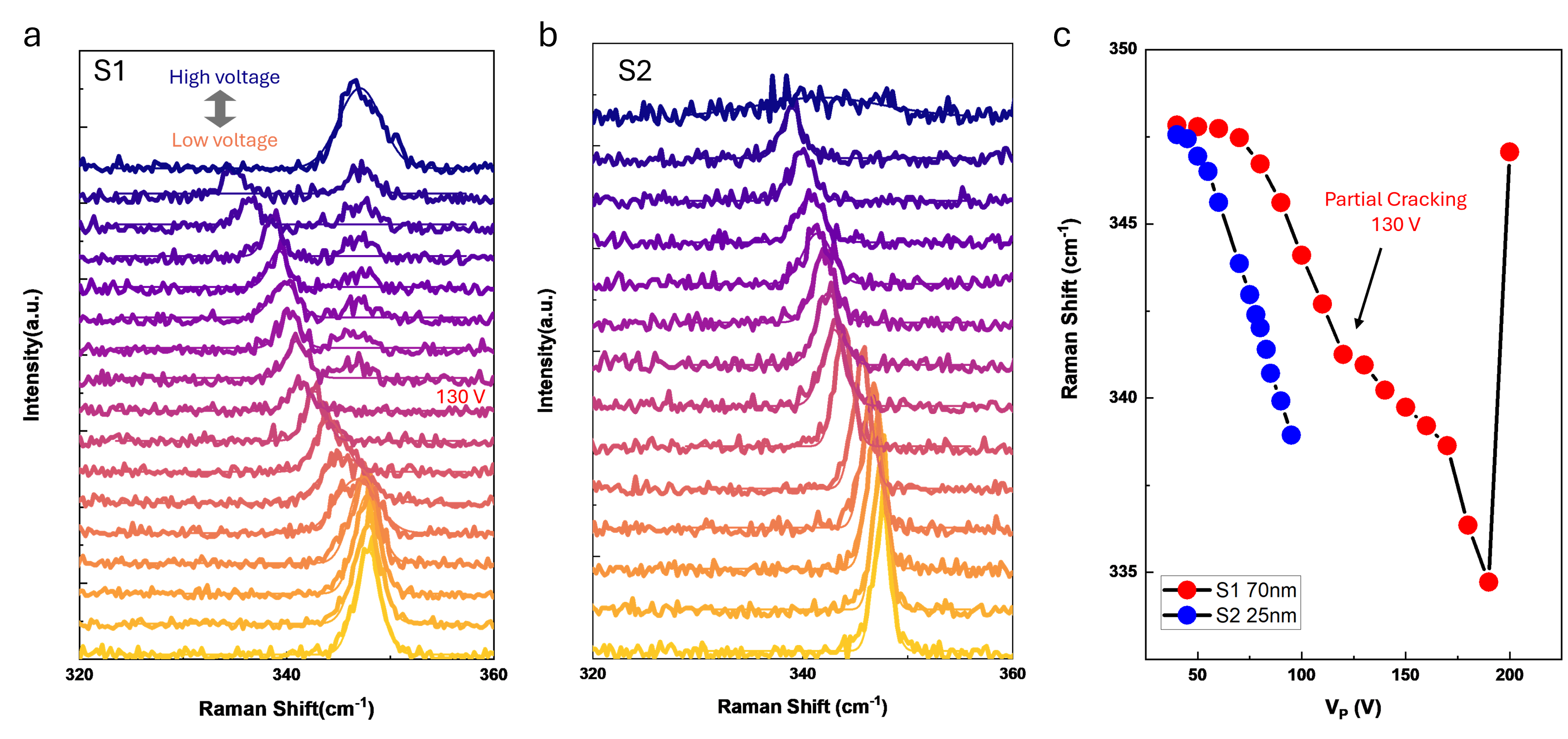}
    \caption{\textbf{Low-temperature strain calibration via Raman spectroscopy.} Evolution of the Raman spectra for two CrSBr samples, (a) S1 (70 nm) and (b) S2 (25 nm), measured at 1.8 K as a function of $V_\textrm{P}$. The emergent peak when $V_\textrm{P}=130$ V is aligned with the unstrained peak position, indicating partial relaxation of the sample. Before this peak emerges, the spectra exhibit a monotonic redshift with increasing $V_\textrm{P}$, indicating a linear response to tensile strain. (c) Extracted Raman peak positions versus $V_\textrm{P}$ for both samples. Note the distinct discontinuity in S1 near 130 V, corresponding to partial cracking and relaxation of the sample.}
\label{CryogenicPerformance}
\end{figure}

\begin{figure}[H]
    \centering
    \includegraphics[width=0.75\linewidth]{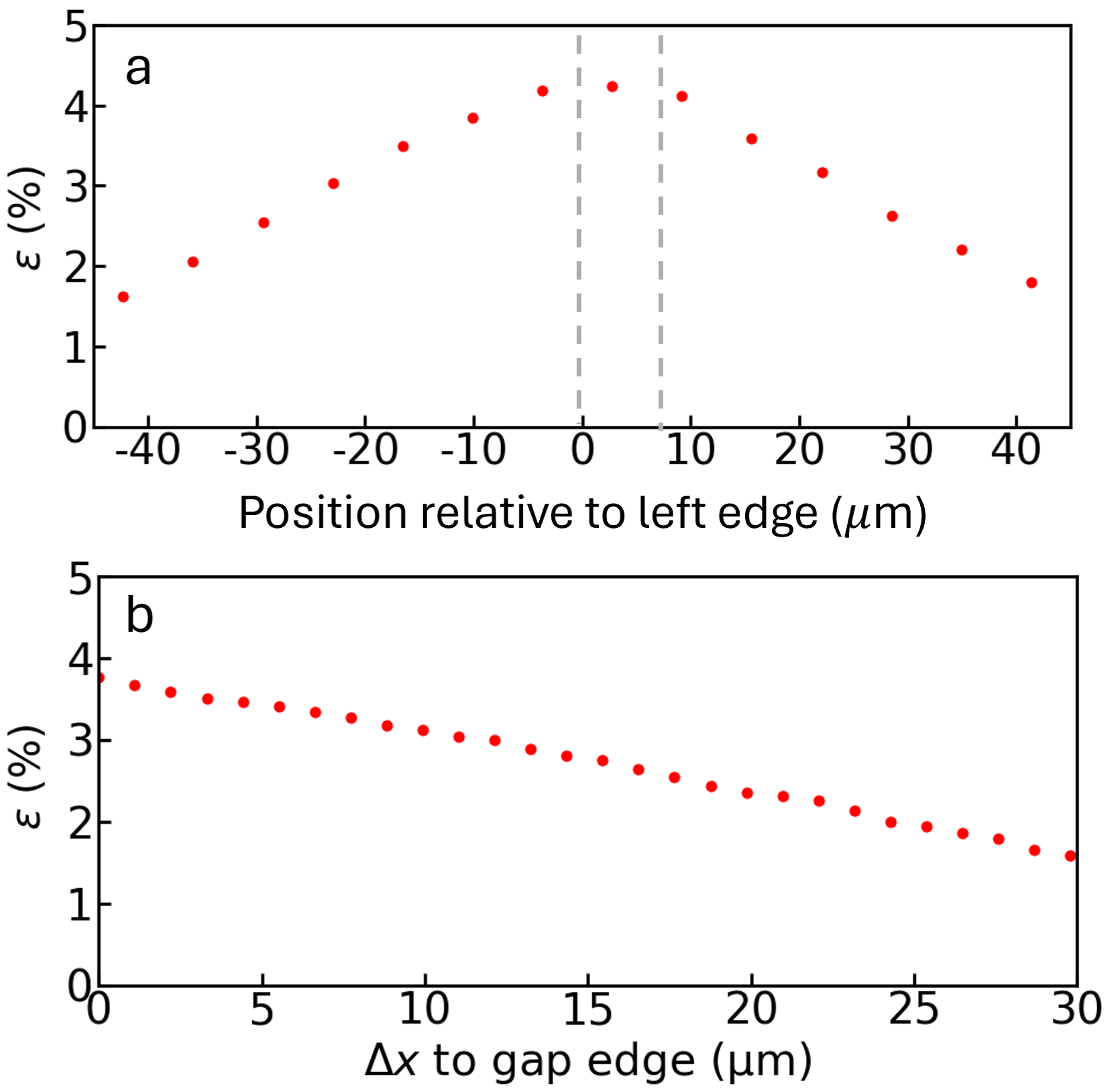}
    \caption{\textbf{Spatial strain distribution profiles.} (a) Strain distribution measured across the gap, showing a symmetric profile with a central strain plateau. The dashed vertical line marks the position of the chip edge.  (b) Linear strain gradient extending over a large distance from the gap edge in the high-strain regime. All strain values were extracted from the Gaussian fitting of spatially resolved Raman spectra.}
\label{Survey}
\end{figure}
\begin{figure}[H]
    \centering
    \includegraphics[width=0.75\linewidth]{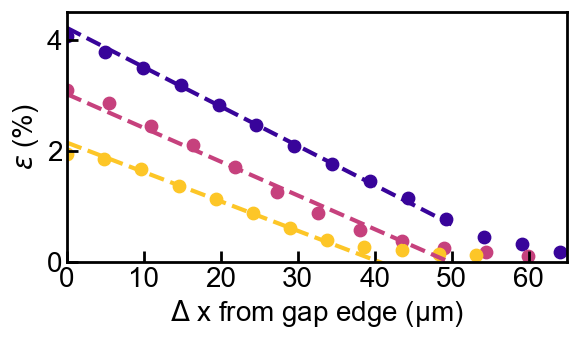}
    \caption{\textbf{Spatial distribution of strain on samples.} Spatial distribution of strain ($\epsilon$) extending from the gap edge for three distinct strain states (4\%, 3\%, 2\%). The dashed lines represent linear fits to the strain decay, demonstrating consistent strain gradient adjustable with fixed strain for material under suspension.}
\label{gradientFit}
\end{figure}

\begin{figure}[H]
    \centering
    \includegraphics[width=0.75\linewidth]{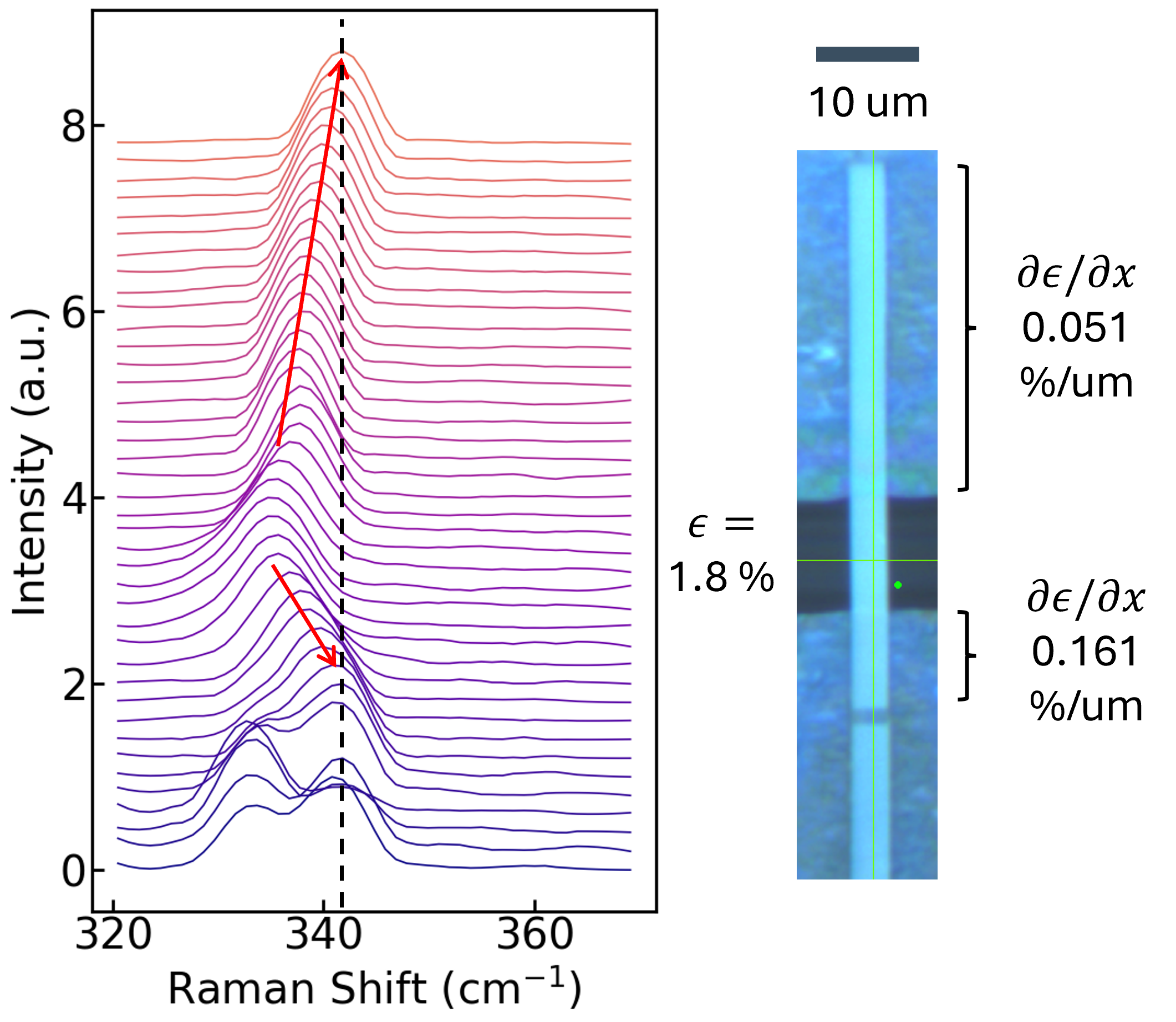}
    \caption{\textbf{Asymmetric strain distribution via different anchoring lengths.} Spatially resolved Raman spectra along a suspended flake with unequal anchoring lengths (top vs. bottom). The varying strain gradients demonstrate that the natural strain relaxation length is not a fundamental lower bound for the minimal length of the sample; rather, shorter anchoring lengths can induce higher strain gradients while maintaining the same applied strain inside the gapped region.}
\label{Asym}
\end{figure}

\begin{figure}[h]
    \centering
    \includegraphics[width=0.6\linewidth]{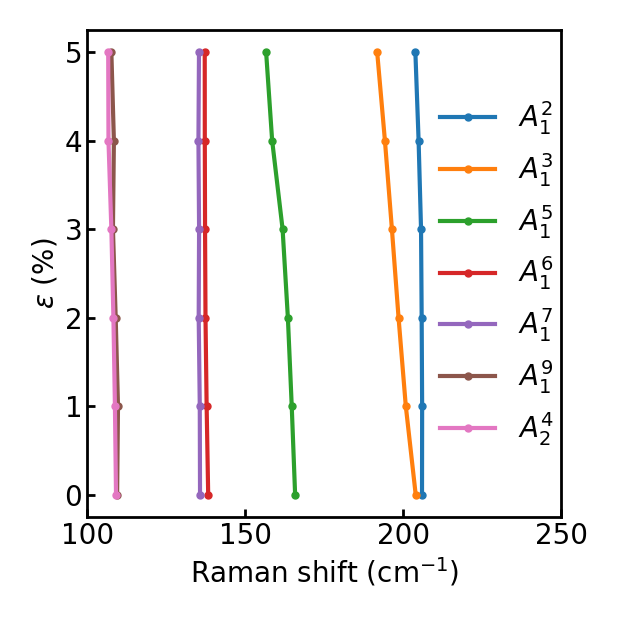}
    \caption{\textbf{DFT-calculated strain dependence of Raman-active phonon modes in $T_d$-WTe$_2$.} Calculated Raman shifts of the experimentally-observed optically active modes as a function of strain applied along the $a$ axis.}
\label{DFTSurvey}
\end{figure}
\begin{figure}[H]
    \centering
    \includegraphics[width=0.6\linewidth]{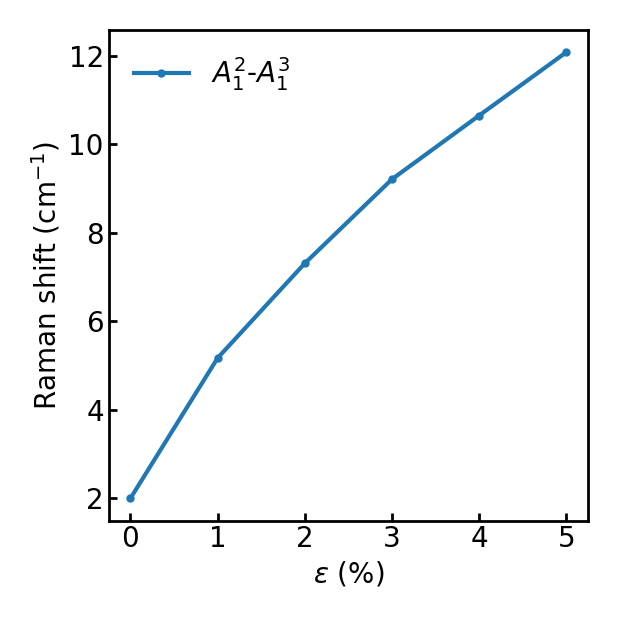}
    \caption{\textbf{DFT-calculated strain-dependent splitting of Raman modes for WTe$_2$.} Calculated wavenumber difference between the \textrm{A$^2_1$} and \textrm{A$^3_1$} modes as a function of strain applied along the $a$ axis.}
\label{DFTDifference}
\end{figure}
\begin{table}[H]
\centering
\begin{tabular}{cccccc}
\hline
Idx & Energy & Irrep & Idx & Energy & Irrep \\
\hline
1  & -0.00  & $B_1$ & 19 & 130.13 & $B_1$ \\
2  &  0.00  & $A_1$ & 20 & 133.89 & $B_1$ \\
3  &  0.00  & $B_2$ & 21 & 135.69 & $A_1$ \\
4  & 11.62  & $A_1$ & 22 & 138.27 & $A_1$ \\
5  & 31.41  & $A_2$ & 23 & 149.31 & $A_2$ \\
6  & 38.93  & $B_1$ & 24 & 149.75 & $B_2$ \\
7  & 77.70  & $A_1$ & 25 & 155.91 & $B_1$ \\
8  & 81.40  & $B_2$ & 26 & 156.79 & $A_2$ \\
9  & 85.67  & $A_2$ & 27 & 158.54 & $B_2$ \\
10 & 91.40  & $B_1$ & 28 & 165.76 & $A_1$ \\
11 &109.12  & $A_2$ & 29 & 176.44 & $B_1$ \\
12 &109.47  & $A_1$ & 30 & 178.06 & $A_1$ \\
13 &110.82  & $B_2$ & 31 & 203.89 & $A_1$ \\
14 &114.32  & $A_2$ & 32 & 205.57 & $B_1$ \\
15 &118.43  & $B_2$ & 33 & 205.89 & $A_1$ \\
16 &119.84  & $B_1$ & 34 & 209.13 & $B_1$ \\
17 &124.59  & $B_1$ & 35 & 231.86 & $A_1$ \\
18 &130.06  & $A_1$ & 36 & 233.86 & $B_1$ \\
\hline
\end{tabular}
\caption{Phonon mode energies and symmetry labels for WTe$_2$ at 0\% strain}
\label{tab:modes_symmetry}
\end{table}
\bibliography{reference_Dec.bib}